\documentclass[useAMS,usenatbib,onecolumn]{mn2e}
\usepackage{natbib}
\usepackage{amsmath}
\usepackage{amsfonts}
\usepackage{graphicx}
\usepackage{tabularx}
\usepackage{multirow}
\usepackage{lscape}
\usepackage{rotating}
\usepackage{pstricks}
\usepackage{color}
\usepackage{ amssymb }

\newcommand{\delete}{\bgroup\markoverwith{\textcolor{red}{\rule[0.5ex]{2pt}{1pt}}}\ULon}

\title{ Heat capacity of low density neutron matter: from quantum to classical regimes}
\author[A. Pastore, N. Chamel and J. Margueron]{A. Pastore$^{1}$\thanks{E-mail:
apastore@ulb.ac.be}, N. Chamel$^{1}$ and J. Margueron$^{2}$\\
$^{1}$Institut d'Astronomie et d'Astrophysique, CP 226, Universit\'e Libre de Bruxelles, B-1050 Bruxelles, Belgium\\
$^{2}$Universit\'e de Lyon, F-69003 Lyon, France; Universit\'e Lyon 1,
             43 Bd. du 11 Novembre 1918, F-69622 Villeurbanne cedex, France\\
             CNRS-IN2P3, UMR 5822, Institut de Physique Nucl{\'e}aire de Lyon}
             
\begin{document}

\date{\today}

\pagerange{\pageref{firstpage}--\pageref{lastpage}} \pubyear{2014}

\maketitle

\label{firstpage}

\begin{abstract}
The heat capacity of neutron matter is studied over the range of densities and temperatures prevailing in neutron-star crusts, 
allowing for the transition to a superfluid phase at temperatures below some critical temperature $T_{sf}$ and including the 
transition to the classical limit. 
Finite temperature Hartree-Fock-Bogoliubov equations (FTHFB) are solved and compared to 
existing approximate expressions.
In particular, the 
formula given by Levenfish and Yakovlev is found to reproduce the numerical results with a high degree of accuracy for temperatures 
$T\leq T_{sf}$. 
In the non-superfluid phase, $T\geq T_{sf}$, the linear approximation 
is valid only at temperature $T\ll T_{{\rm F} n}$ ($T_{{\rm F} n}$ being the Fermi temperature of the neutron gas)
which is rarely the case in the shallow layers of the neutron star's crust.
A non-perturbative interpolation between the quantal and the classical regimes is proposed here. 
The heat capacity, conveniently parametrized solely in terms of $T_{sf}$, $T_{{\rm F} n}$, and the neutron number density $n_n$, can be easily 
implemented in neutron-star cooling simulations. 

\end{abstract}


\section{Introduction}

Neutron stars (NSs) are formed from the catastrophic gravitational core collapse of stars with a mass $M> 8 M_\odot$ \citep{Book:Haensel2007}. 
During the first tens of seconds after the collapse, the newly formed proto-neutron star with a radius of about 50 km stays very hot 
with internal temperatures of the order of  $\sim 10^{11}-10^{12}$~K. But it rapidly cools down and shrinks into an ordinary neutron 
star by emitting neutrinos. As the temperature drops, the outer layers of the star crystallize into a solid crust, whose innermost 
region is permeated by ``free'' neutrons \citep[see][for a review of neutron-star crusts]{Chamel2008a}. At temperatures $T$ lower 
than the Fermi temperature $T_{{\rm F}n}$, the neutron liquid becomes degenerate. As the temperature reaches some critical 
temperature $T_{sf}$, the neutron liquid undergoes a transition to a spin-singlet superfluid phase by forming Cooper 
pairs, like electrons in conventional superconductors. Due to its relatively low neutrino emissivity, the crust of a newly-born neutron star cools 
less rapidly than the core and thus stays hotter. As a result, the surface temperature decreases slowly during 
the first ten to hundred years and then suddenly drops  when the \textsl{cooling wave} from the core reaches the surface. 
The evolution of the surface temperature of a young neutron star thus depends essentially on the thermal properties of its 
crust~\citep{Lattimer1994a, Gnedin2001}. 

While the cooling of very young neutron stars has not been observed yet, the thermal relaxation of neutron-star crusts has been 
recently monitored in a few quasi-persistent soft X-ray transients~\citep[see][for a recent review]{Page2012:book}. In these binary systems, 
usually called Low-Mass X-Ray Transients (LMXRT),
a neutron star accretes matter from a companion star during severals years or decades, driving the neutron-star crust out of its thermal 
equilibrium with the core. After the accretion stops, the heated crust relaxes towards equilibrium. 
While the thermal relaxation of young isolated NSs is mostly governed by the thermal properties of the densest layers of the NS crust, 
the thermal relaxation of LMXRT is expected to be very sensitive to the thermal properties of the shallower layers of the crust, especially near 
the transition between the inner and the outer crusts~\citep{Page2012:book}. In this region of the NS crust, $T$ may be not only higher that $T_{sf}$ 
but also to $T_{{\rm F}n}$ so that neutrons can be treated as a classical gas: their heat capacity does not increase linearly with $T$ as in the deeper 
crustal regions where $T_{sf}\leq T\ll T_{{\rm F}n}$, but becomes essentially independent of $T$. Assuming that neutrons are degenerate in all 
regions of the crust may thus considerably overestimates the crustal heat capacity hence also the thermal relaxation time of LMXRT.

In this paper, we study the neutron contribution to the heat capacity of the neutron-star crust focusing on the region close to the neutron drip 
transition. Although the presence of inhomogeneities may alter the neutron heat capacity~\citep[see, e.g., ][]{Book:Margueron2012}, we shall simply consider 
here that the neutron liquid is homogeneous, as generally assumed in neutron-star cooling simulations~\citep[see, e.g., ][]{Page2012:book}. 
Over the past years, the properties of dilute neutron matter have been extensively studied. Microscopic calculations based on realistic 
nucleon-nucleon potentials and following different many-body approaches tend to yield similar results, at least at low enough densities~\citep{Baldo2012}. 
On the other hand, most of these calculations have been restricted to $T=0$, and therefore they cannot be directly applied to neutron-star 
cooling simulations. For this reason, we have employed the finite-temperature Hartree-Fock-Bogoliubov (FTHFB) method, which is briefly 
reviewed in Section~\ref{sec:FTHFB}. The neutron heat capacity is studied in Section~\ref{sec:heat-cap} for both the classical and 
degenerate regimes.

\section{Finite-temperature Hartree-Fock-Bogoliubov method}
\label{sec:FTHFB}

The neutron pairing phenomenon is studied here using the self-consistent finite-temperature Hartree-Fock-Bogoliubov (FTHFB) method with 
effective nucleon-nucleon interactions~\citep{Goodman1981a,Sandulescu2004b,Book:Margueron2012,Pastore2012}. This method not only provides 
a convenient parametrization of microscopic neutron-matter calculations based on realistic potentials at $T=0$, but can also be applied to 
determine in a consistent way the properties of neutron matter at finite temperatures including the transition between the quantum 
and classical regimes. Moreover, the same method can also be applied to study inhomogeneous nuclear matter like neutron-star 
crusts~\citep[see, e.g., ][]{Book:Margueron2012}.

A key quantity for characterizing the superfluid phase is the pairing gap function $\Delta_n(k)$, $k$ being the wave vector. This function, 
which is related to the binding energy of a pair, depends also on the temperature $T$ ($\Delta_n$ vanishing at $T\geq T_{sf}$) and on the neutron 
number density $n_n$. The pairing gap can be determined by solving the following equations (the units used throughout this paper are such that 
the Boltzmann constant is $k_{\rm B}=1$): 
\begin{equation}
\label{bcs:pnm1}
\Delta_{n}(k)=-\frac{1}{4} \int \frac{d^{3}k^\prime}{(2\pi)^{3}}v(k, k^\prime)\frac{\Delta_{n}(k^\prime)}{E_{n}(k^\prime)}\tanh \left(\frac{E_{n}(k^\prime)}{2T} \right),
\end{equation}
\begin{equation}
\label{bcs:pnm2}
n_n=\int \frac{d^{3}k^\prime}{(2\pi)^{3}}\left[1-\frac{\varepsilon_{n}(k^\prime)-\mu_{n}}{E_{n}(k^\prime)} \tanh \left(\frac{E_{n}(k^\prime)}{2 T} \right) \right],
\end{equation}
where $\varepsilon_n(k)$ denotes the neutron single-particle energies, $E_{n}(k)=\sqrt{(\varepsilon_{n}(k)-\mu_{n})^{2}+\Delta_{n}(k)^{2}}$ 
the neutron quasi-particle energies, $\mu_n$ the neutron chemical potential, and $v(k, k^\prime)$ the matrix elements of the effective pairing 
interaction. As in conventional superconductors, the effective pairing interaction may not necessarily be the same as the effective interaction 
which determines the single-particle properties in dense matter. For the latter, we use a zero-range effective interaction of the Skyrme type~\citep{Skyrme1958,Bender2003a}
\begin{eqnarray}\label{eq:skyrme}
v(\pmb{r}_{i},\pmb{r}_{j}) & = & t_0(1+x_0 P_\sigma)\delta({\pmb{r}_{ij}})+\frac{1}{2} t_1(1+x_1 P_\sigma)\frac{1}{\hbar^2}\left[p_{ij}^2\,
\delta({\pmb{r}_{ij}}) +\delta({\pmb{r}_{ij}})\, p_{ij}^2 \right] +t_2(1+x_2 P_\sigma)\frac{1}{\hbar^2}\pmb{p}_{ij}\cdot\delta(\pmb{r}_{ij})\,\pmb{p}_{ij}\nonumber \\
&&+\frac{1}{6}t_3(1+x_3 P_\sigma)n_n(\pmb{r})^\alpha\,\delta(\pmb{r}_{ij})+\frac{\rm i}{\hbar^2}W_0(\pmb{\hat\sigma_i}+\pmb{\hat\sigma_j})\cdot
\pmb{p}_{ij}\times\delta(\pmb{r}_{ij})\,\pmb{p}_{ij} \quad , 
\end{eqnarray}
where $\pmb{r}_{ij} = \pmb{r}_i - \pmb{r}_j$, $\pmb{r} = (\pmb{r}_i + 
\pmb{r}_j)/2$, $\pmb{p}_{ij} = - {\rm i}\hbar(\pmb{\nabla}_i-\pmb{\nabla}_j)/2$
is the relative momentum, $\pmb{\hat\sigma_i}$ and $\pmb{\hat\sigma_j}$ are Pauli spin matrices, and $P_\sigma$ is the two-body spin-exchange 
operator. Skyrme effective interactions were shown to be well suited for reproducing microscopic calculations at both zero and finite 
temperatures~\citep{Fantina2012}. Using Eq.~(\ref{eq:skyrme}), the neutron single-particle energies can be expressed as
\begin{equation}
\varepsilon_n(k)=\frac{\hbar^{2}k^{2}}{2m_n^*}+U_n,
\end{equation}
where $U_n$ denotes the (self-consistent) mean potential field felt by neutrons whereas $m_n^*$ is the (self-consistent) mean neutron effective 
mass given by~\citep{Bender2003a}
\begin{equation}
\frac{\hbar^{2}}{2m_n^{*}}=\frac{\hbar^{2}}{2m_n}+\frac{1}{8}\left[t_1(1-x_1)+3 t_2(1+x_2)\right]n_n, 
\end{equation}
$m_n$ being the ``bare'' neutron mass. 
Since the FTHFB equations~(\ref{bcs:pnm1})-(\ref{bcs:pnm2}) depend only on the difference $\varepsilon_{n}(k)-\mu_{n}$, the pairing gap function 
is actually 
independent of the potential $U_n$. It is therefore convenient to introduce a shifted chemical potential $\nu_n\equiv\mu_n-U_n$. In this way, 
we have $\varepsilon_n(k)-\mu_{n}=\epsilon_n(k)-\nu_n$ with  
\begin{equation}
\epsilon_n(k)=\frac{\hbar^{2}k^{2}}{2m_n^*},
\label{eq:shiftedenergy}
\end{equation}
so that $U_n$ disappears from the FTHFB equations. 

As for pairing, we shall also consider zero-range effective interaction of the form
\begin{equation}\label{func:garrido}
v(\pmb{r}_{i},\pmb{r}_{j})=v^{\pi}(n_n(\pmb{r}))\delta(\pmb{r}_{ij}) \quad .
\end{equation}
In this case, the matrix elements of the pairing interaction reduce to $v(k, k^\prime)=v^\pi(n_n)$. It can thus be seen from Eq.~(\ref{bcs:pnm1}) 
that the pairing gap function is independent of $\pmb{k}$. As shown by \cite{Chamel2010}, the pairing strength $v^{\pi}(n_n)$ can be directly inferred 
from microscopic calculations of the pairing gap function $\Delta_n(n_n)$ at $T=0$ as follows
\begin{equation}\label{func:bsk16}
v^\pi(n_n)=-\frac{8\pi^2}{\sqrt{\nu_n}}\left(\frac{\hbar^2}{2 m_n^*}\right)^{3/2} \biggl[ 2\log\left(\frac{2\nu_n}{\Delta_n}\right)+ \Lambda\left(\frac{\varepsilon_\Lambda}{\nu_n}\right) \biggr]^{-1}
\end{equation}
with
\begin{equation}
\label{eq.12}
\Lambda(x)=\log (16 x) + 2\sqrt{1+x}-2\log\left(1+\sqrt{1+x}\right)-4\, .
\end{equation}
The only free parameter is the pairing cutoff $\varepsilon_\Lambda$ which needs to be introduced to regularize divergences in Eq.~(\ref{bcs:pnm1}) 
arising from the zero-range of the interaction. The pairing functional defined in this way ensures that the solution of the FTHFB 
equations~(\ref{bcs:pnm1})-(\ref{bcs:pnm2}) coincides exactly with $\Delta_n$ at $T=0$. With the effective interactions considered here, 
the gap equations (\ref{bcs:pnm1}) and (\ref{bcs:pnm2}) 
for $T<T_{sf}$ reduce to 
\begin{equation}
\label{bcs_red:pnm1}
1=-\frac{1}{8} v^\pi(n_n) \int_{-\infty}^{\nu_n+\varepsilon_\Lambda} d\epsilon  \frac{D(\epsilon)}{E(\epsilon)}\tanh \left(\frac{E(\epsilon)}{2T} \right)\, ,
\end{equation}
\begin{equation}
\label{bcs_red:pnm2}
n_n=\int_{-\infty}^{+\infty} d\epsilon D(\epsilon) \left[1-\frac{\epsilon-\nu_{n}}{E(\epsilon)} \tanh \left(\frac{E(\epsilon)}{2 T} \right) \right]\, ,
\end{equation}
where $E(\epsilon)=\sqrt{\Delta_n^2 + (\epsilon-\nu_n)^2}$, and $D(\epsilon)$ denotes the density of neutron single-particle states, defined as 
\begin{equation}
 D(\epsilon)=2 \int \frac{d^{3}k}{(2\pi)^{3}} \delta(\epsilon-\epsilon_n(k))=\frac{1}{2\pi^2}\left(\frac{2 m_n^*}{\hbar^2}\right)^{3/2}\sqrt{\epsilon}\, .
\end{equation}

\section{Heat capacity of dilute neutron matter}
\label{sec:heat-cap}

The neutron heat capacity $C_{V}^{n}$ (per unit volume) is defined by~\citep{Book:LL1}
\begin{equation}\label{cv}
C_{V}^{n}(n_n,T)=T\left.\frac{\partial S^{n}}{\partial T}\right|_{n_n} ,
\end{equation}
where $S^{n}$ is the entropy density given by 
\begin{equation}\label{entropy}
S^{n}(n_n,T)=-\int d\epsilon \left[f(\epsilon)\log f(\epsilon) +(1-f(\epsilon))\log(1-f(\epsilon))\right]\, ,
\end{equation}
with $f(\epsilon)=\left(1+\exp(E(\epsilon)/T)\right)^{-1}$\, .
Injecting Eq.~(\ref{entropy}) into (\ref{cv}) yields the general expression
\begin{equation}
\label{eq:cv}
C_{V}^{n}(n_n,T)=\int d\epsilon (1-f(\epsilon))f(\epsilon) \left[ \left( \frac{E(\epsilon)}{T}\right)^{2}+\frac{\epsilon-\nu_n}{T}\frac{\partial \nu_n}{\partial T}
-\frac{\Delta_n}{T}\frac{\partial \Delta_n}{\partial T}\right]\, .
\end{equation}

At high enough density, as in the core of a neutron star, neutron pairs are very loosely bound so that $\Delta_n\ll \nu_n$. In the weak 
coupling approximation~\citep{Book:LL2}, Eq.~(\ref{bcs_red:pnm1}) 
is replaced by  
\begin{equation}
\label{bcs_weak_coupling}
1\approx -\frac{1}{8} v^\pi(n_n)D(\nu_n) \int_{-\infty}^{\nu_n+\varepsilon_\Lambda} d\epsilon  \frac{1}{E(\epsilon)}\tanh \left(\frac{E(\epsilon)}{2T} \right)\, ,
\end{equation}
and Eq.~(\ref{bcs_red:pnm2}) by $\nu_n\approx T_{{\rm F}n}$, where { the Fermi temperature is defined by $T_{{\rm F}n}=\epsilon_n(k=k_{{\rm F}n})$, see e.g.
Eq.~(\ref{eq:shiftedenergy}), and}
where $k_{{\rm F}n}=(3\pi^2 n_n)^{1/3}$ denotes the Fermi wave number. Equation~(\ref{bcs_weak_coupling})
can be further approximated by~\citep{Book:LL2}
\begin{equation}
\label{bcs_weak_coupling2}
\log\left(\frac{\Delta_n(0)}{\Delta_n(T)}\right)\approx \int_0^{+\infty}\frac{dx}{\sqrt{x^2+u^2}(1+\exp(\sqrt{x^2+u^2})}\, ,
\end{equation}
with $u=\Delta_n(T)/T$ and the integration has been extended to $\pm\infty$. 
In their seminal paper, \cite{Levenfish1994} have solved Eq.~(\ref{bcs_weak_coupling2}) to calculate the 
neutron heat capacity at densities relevant for neutron star cores, and fitted the numerical results with the 
following analytical expression:  
\begin{equation}\label{eq:levenfish}
C_{V}^n(n_n,T)=R_{00}(u) C_{V}^{n\,(nor)}(n_n,T)\,,
\end{equation}
where $C_{V}^{n\,(nor)}$ is the heat capacity of non-superfluid neutrons, which in the quantum regime of strongly degenerate 
neutrons, $T\ll T_{{\rm F}n}$, is approximately given by~\citep{Book:LL2}
\begin{equation}\label{cv:lin}
C_{V}^{n\,(FG)}(n_n,T)=\frac{\hbar^{2}}{4m_n^{*}} n_n^{1/3} \left( \frac{\pi}{3}\right)^{2/3} T=\frac{\pi^2}{2}n_n\frac{T}{T_{{\rm F}n}} \,.
\end{equation}
In Eq.~(\ref{eq:levenfish}), $R_{00}$ is a correction factor introduced to account for the effects of superfluidity
and was parametrized as~\citep{Levenfish1994} 
\begin{equation}\label{eq:r00}
R_{00}(u)=\left[ a_{0}+\sqrt{(a_{1})^{2}+(a_{2} u)^{2}}\right]^{\gamma}\exp\left( b_{0}-\sqrt{b_{1}^{2}+u^{2}}\right)\;,
\end{equation}
where $u=\sqrt{1-\tau}\left( c_{0}-c_{1}/\sqrt{\tau}+c_{2}/\tau\right)$, $\tau=T/T_{sf}$. 

Equations~(\ref{eq:levenfish}), (\ref{cv:lin}), (\ref{eq:r00}) have been widely applied in neutron-star cooling simulations. 
However, the weak-coupling approximation used to obtain these expressions may not be very accurate for the dilute neutron 
liquid that permeates the inner crust of a neutron star, especially in the shallowest region where $\Delta_n$ becomes 
comparable to $\nu_n$. Note that in this case, the chemical potential $\nu_n$ may be very different from the Fermi energy.
{ It was found, for instance, that the low density regime could be located in the BCS-BEC crossover, either close to the
unitary limit, or, in some cases, almost entering the BEC regime (where the chemical potential becomes negative as $T\rightarrow0$)
~\citep{Margueron2007c}. 
The FTHFB equations (\ref{bcs_red:pnm1})-(\ref{bcs_red:pnm2}) are however able to describe both the weak and the strong 
coupling regimes~\citep{Nozieres1985}.}
For this reason, we have preferred to solve the set of coupled equations (\ref{bcs_red:pnm1}) 
and (\ref{bcs_red:pnm2}) without any further approximation { instead of Eq.~(\ref{bcs_weak_coupling2})}. 
The predictions for the heat capacity in the nonsuperfluid and superfluid phases will be 
discussed separately in Sections~\ref{subsec:normal} and \ref{subsec:super} respectively. 

\begin{figure}
\begin{center}
\includegraphics[width=0.38\textwidth,angle=-90]{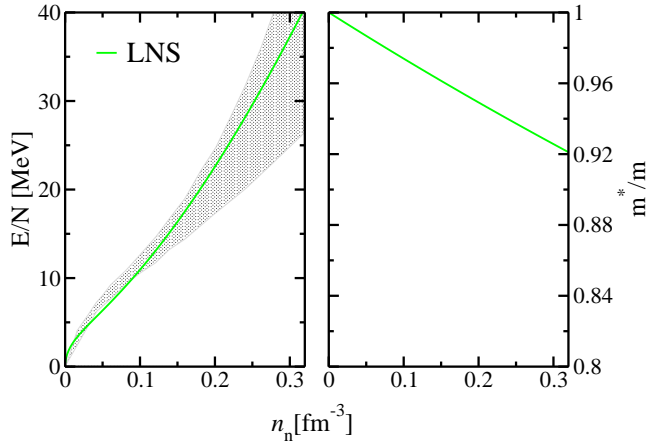}
\end{center}
\caption{(Color online) Equation of state (left panel) and effective mass (right panel) in neutron matter for the LNS Skyrme functional~\citep{Cao2006}. The grey area 
represents the range of equations of state obtained from quantum Monte Carlo calculations~\citep{Gandolfi2012}. See text for details.}
\label{eos:pnm}
\end{figure}

\begin{figure}
\begin{center}
\includegraphics[width=0.38\textwidth,angle=-90]{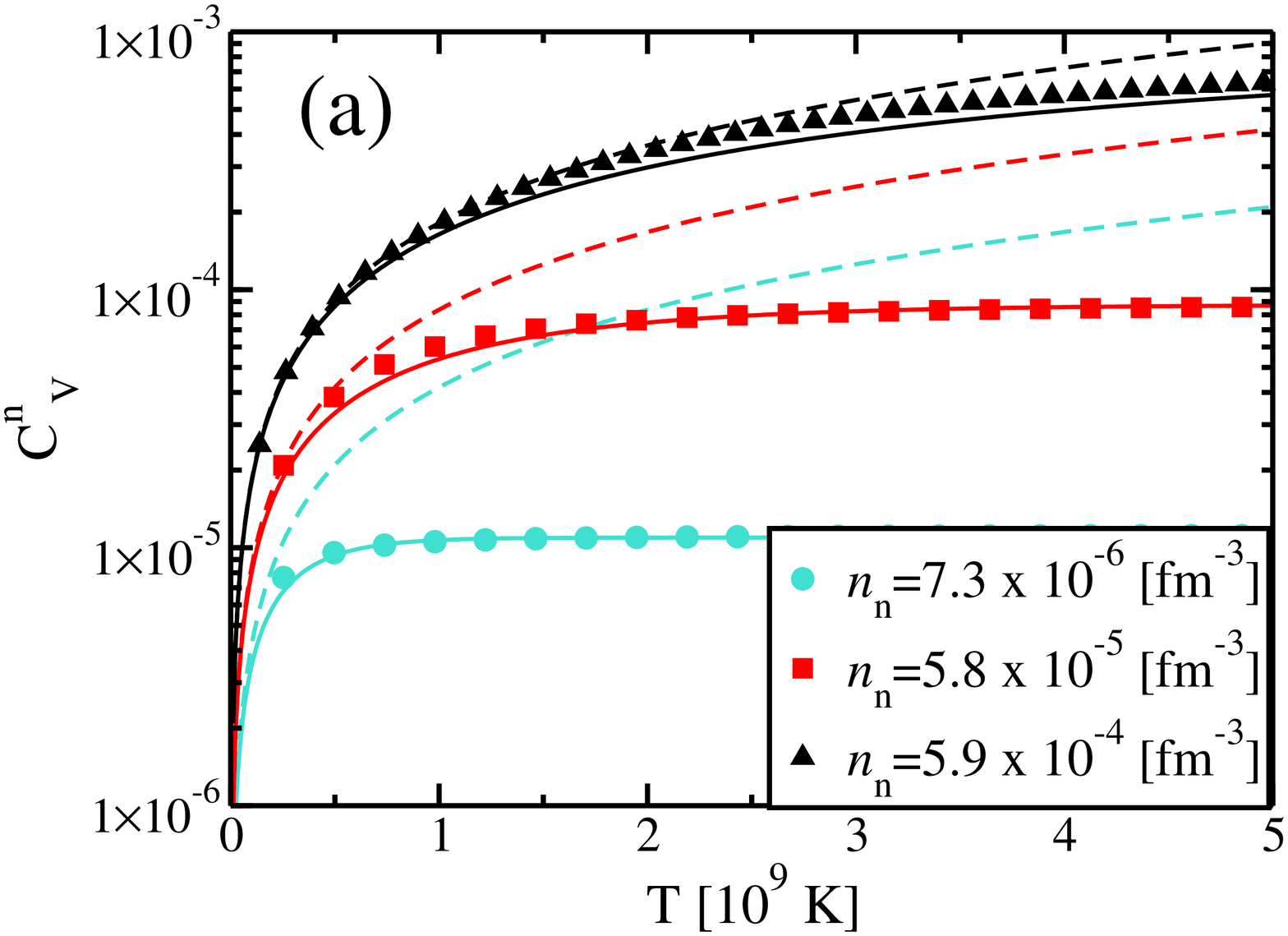}
\includegraphics[width=0.38\textwidth,angle=-90]{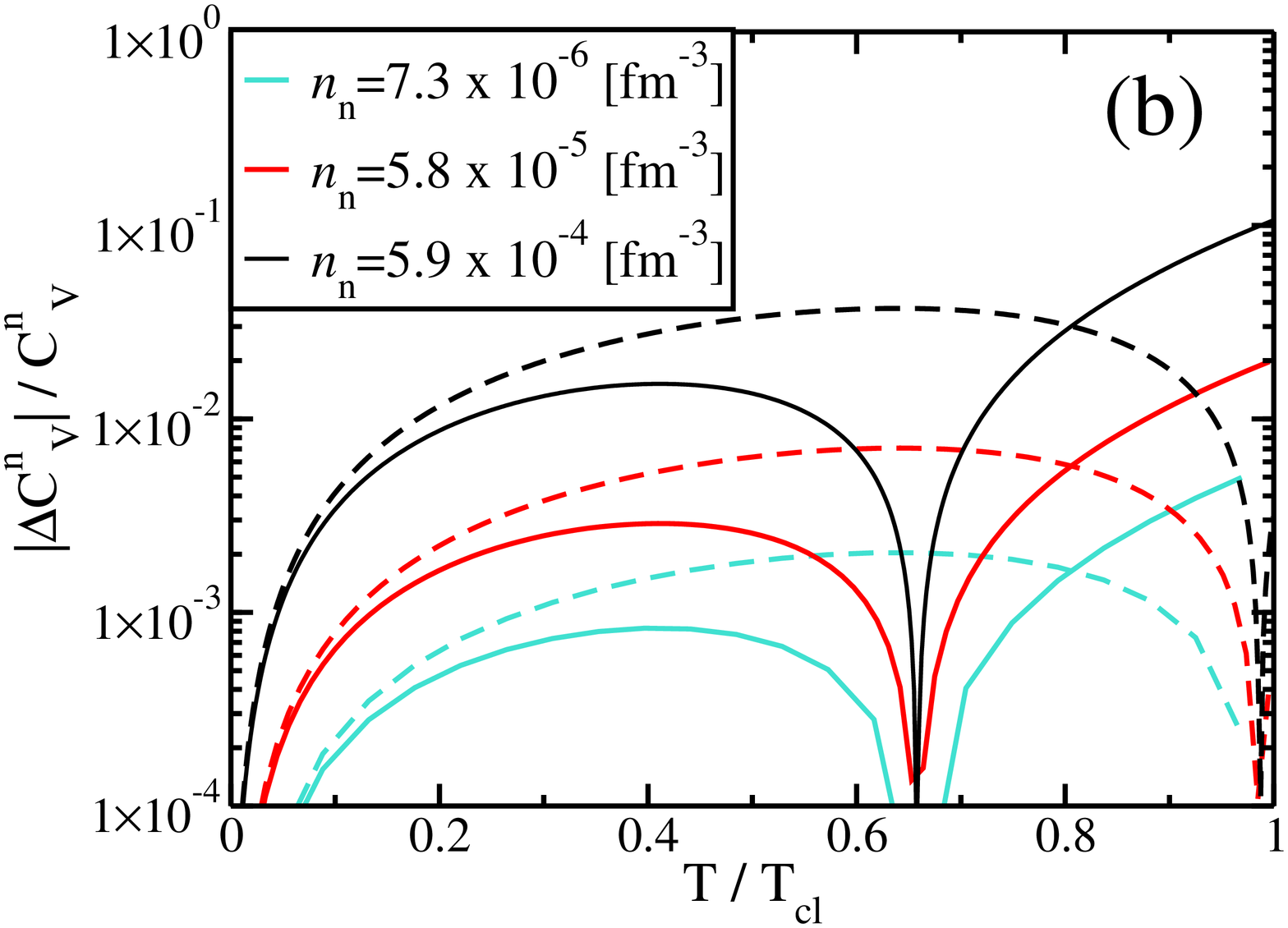}
\end{center}
\caption{(Color online) On the left panel (a): heat capacity of nonsuperfluid neutrons as a function of temperature for different values of the density 
comparing the exact results (symbols) with the linear approximation given by Eq.~(\ref{cv:lin}) (dashed lines) and with the 
improved analytical formula given in Eq.~(\ref{cv:exp}) (solid lines). On the right panel (b): the correction to the heat capacity induced dy the exchange of long-wavelength phonons: solid lines have been calculated using $x_{c}=0.2$, while dashed with $x_{c}=0.3$. See text for details.}
\label{cvn:hf}
\end{figure}

\subsection{Nonsuperfluid phase: from the classical to the degenerate regimes}
\label{subsec:normal}

In this section, we first study the heat capacity of a non-superfluid neutron liquid. For this purpose, we have considered the LNS
Skyrme functional~\citep{Cao2006}, which was fitted to many-body calculations based on the Brueckner method and using realistic 
two- and three-body forces. In particular, the parameters of this functional were adjusted so as to reproduce both the equation of state and the 
effective mass in neutron matter as obtained from many-body calculations. As shown in Fig.~\ref{eos:pnm}, the equation of state given by 
this functional is in good agreement with more recent quantum Monte Carlo calculations~\citep{Gandolfi2012}.

In Fig.\ref{cvn:hf} (a), the neutron heat capacity calculated from Eq.~(\ref{eq:cv}) with $\Delta_n=0$ is compared with the results obtained using the 
linear approximation~(\ref{cv:lin}). Large deviations can be observed in the low density region 
where $n_n\lesssim 10^{-3}$~fm$^{-3}$ or equivalently $\rho\lesssim 10^{12}$~g~cm$^{-3}$ \citep[see, e.g.,][]{Pearson2012}. 
At high temperatures $T\gg T_{{\rm F}n}$, the heat capacity tends to the classical limit~\citep{Book:LL1}
\begin{equation}\label{cv:classical}
C_{V}^{n\,(cl.)}(n_n)=\frac{3}{2} n_n\, .
\end{equation}

{\color{red} In the $^3$He normal phase, the coupling to phonons is known to largely correct Eq.~(\ref{cv:lin}) by the following term
\begin{equation}
\Delta C_V^{n(FG)}(n_n,T)=-\frac{3}{20}\pi^4 n_n B \left(\frac{T}{T_F}\right)^3 \log \frac{T}{T_{c}},
\end{equation}
where
\begin{equation}
B\approx -\frac 1 2 \left[ A_{0s}^2\left(1-\frac{\pi^2}{12}A_{0s}\right)+3A_{0a}^2\left(1-\frac{\pi^2}{12}A_{0a}\right)\right]
\label{eq:factorB}
\end{equation}
and $A_{0s}=F_0/(1+F_0)$ and $A_{0s}=G_0/(1+G_0)$, $F_0$ (resp. $G_0$) being the density density (resp. spin-density spin-density) 
Landau parameter~\citep{baym2008landau} and $T_{c}=x_{c}T_{F}$ is a cut-off temperature.
Eq.~(\ref{eq:factorB}) is the lowest order approximation considering only the monopolar contribution.
Highest order contributions have been calculated by \cite{Pethick1973}.
The correction $\Delta C_V^{n(FG)}$ normalized to $C_V^{n(FG)}$ is shown in Fig.\ref{cvn:hf} (b) for the three densities (dashed-dotted lines) 
and for two different choices of the parameter $x_{c}$ \citep[see, e.g.,][]{baym2008landau}. 
This correction is shown up to the classical temperature $T_{cl}$ since beyond this temperature, the  system could not be treated with the
same quantal approximation methods, and the limit is given by the classical expression~(\ref{cv:classical}).
It is a negligible correction in the case of very diluted neutron matter (less than 1\% in most of the cases). 
This is mainly due to the absolute value of the Landau Parameters $F_0$ and $G_0$ which are small compared to 1, for the considered
densities.
}

\begin{figure}
\begin{center}
\includegraphics[width=0.4\textwidth,angle=-90]{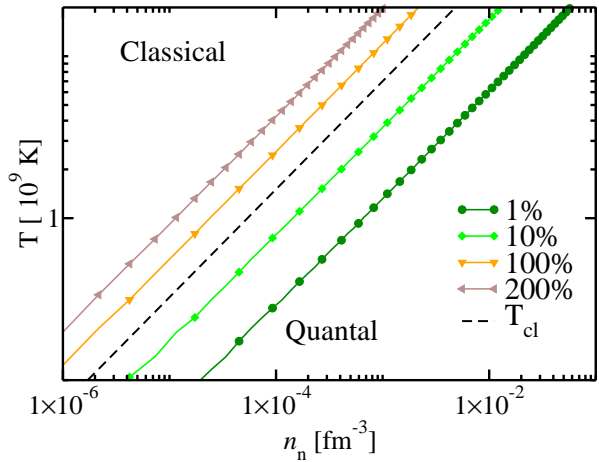}
\end{center}
\caption{(Color online) Temperature as a function of density corresponding to fixed values of the error function 
$\delta(T, n_{n})$. The dashed line represents the characteristic temperature $T_{cl}$ appearing in Eq.~(\ref{cv:exp}) and 
delimiting the classical from the quantum regime. See text for details.}
\label{cvn:err}
\end{figure}

In order to assess more precisely the validity of the linear approximation~(\ref{cv:lin}), we have calculated the relative deviation 
\begin{equation}\label{cv:error}
\delta(T,n_{n})=\frac{|C_{V}^{n}(T,n_{n})-C_{V}^{n\,(FG)}(T,n_{n})|}{C_{V}^{n}(T,n_{n})}\;,
\end{equation}
for the range of neutron densities encountered in neutron-star crusts~\citep[see, e.g.,][]{Pearson2012}. 
Lines of constant $\delta$ in a temperature-density diagram are plotted in Fig.~\ref{cvn:err}. Whereas $\delta$ is of the order of a few percents 
or less at the bottom of the neutron-star crust for temperatures $T\la 10^{10}$~K, errors can be much larger in the shallower layers of the crust near the neutron-drip point where $n_n\rightarrow 0$. 

The exact results for the neutron heat capacity can be represented by the following simple analytical expression 
\begin{equation}\label{cv:exp}
C_{V}^{n\,(FG-cl)}(n_n,T)\approx \frac{3}{2}n_{n} \biggl[ 1-\exp\left(-\frac{T}{T_{cl}}\right)\biggr]
\end{equation}
where $T_{cl}\equiv 3 T_{{\rm F}n}/\pi^2$ is a characteristic temperature delimiting the classical and quantum regimes.{\color{blue}
This expression yields the correct asymptotic behaviours: at low temperatures ($T/T_{cl}\ll1$) Eq.(\ref{cv:exp}) reduces to Eq.(\ref{cv:lin}), while at high temperatures ($T/T_{cl}\gg1$), Eq.(\ref{cv:exp}) tends to the classical limit (\ref{cv:classical}). }
%We notice that this equation gives the correct  asymptotic behaviors: for $T/T_{cl}\ll1$ we obtain the linear approximation given in Eq.\ref{cv:lin},. while for  $T/T_{cl}\gg1$ we obtain the classical limit. It is worth noticing that these two limits do not commute.   } 
As shown in Fig. \ref{cvn:hf}, 
Eq.~(\ref{cv:exp}) provides a very accurate interpolation between the low and high temperature limits. 
The classical temperature $T_{cl}$ is also represented in Fig.~\ref{cvn:err}, showing that at $T\approx T_{cl}$, the relative deviation
$\delta(T,n_{n})$ amounts to $\sim 30$\%.

Note that in Eq.~(\ref{cv:exp}), the effect of the neutron-neutron interactions are embedded in the neutron effective mass appearing 
in the expression of the Fermi temperature $T_{{\rm F}n}$. However, $m_n^* \approx m_n$ at densities below $n_n\sim 10^{-2}$~fm$^{-3}$ therefore the 
expression~(\ref{cv:exp}) is almost independent of the choice of the functional over this range of densities.

While in this paper, we are mostly interested in neutron matter, Eq.~(\ref{cv:exp}) can be easily 
adapted to evaluate the heat capacity of any other fermion species, e.g. electrons.

\subsection{Superfluid phase}
\label{subsec:super}

\begin{figure}
\begin{center}
\includegraphics[width=0.38\textwidth,angle=-90]{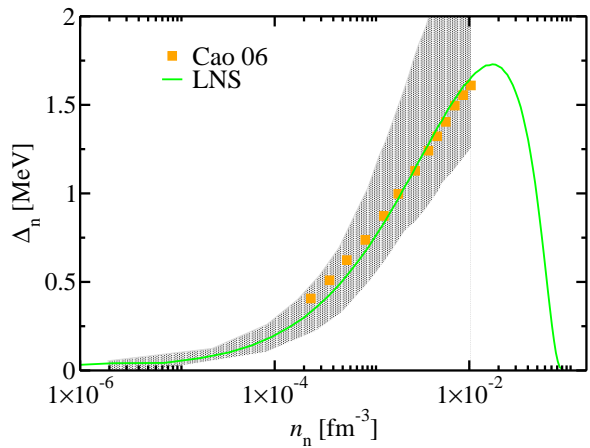}
\end{center}
\caption{(Color online)  $^1\text{S}_{0}$ neutron pairing gap as a function of density as obtained from Eq.~(\ref{fit:eq}) (solid line)
and from \citep{Cao2006a} (symbols). The gray band is extracted from  taking the most recent $ab-initio$ results of Ref.~\citep{gezerlis2010}.}
\label{gap:pnm:dddi:A}
\end{figure}

Let us now study the heat capacity in the superfluid regime and the transition to the normal phase at $T=T_{sf}$. 
In our calculations, we use an effective pairing interaction which is adjusted to mimic more microscopic interactions based on Brueckner method~\citep{bru55}.
We have adopted the ones obtained by \cite{Cao2006a} which includes medium polarization effects beyond the mean-field. 
The reason for calibrating our effective pairing interaction to these results is that they were determined using the same Brueckner approach
as the one we have used to fix the mean-field and the effective mass, see Section~\ref{subsec:normal}. In this way, the particle-hole and the 
particle-particle channels are consistently defined.

The microscopic pairing gaps calculated by \cite{Cao2006a} at $T=0$ can be conveniently parametrized by the following expression~\citep{Kaminker2001}
\begin{equation}\label{fit:eq}
\Delta_n(T=0)=\Theta(k_{max}-k_{Fn})\Delta_{0}\frac{k_{Fn}^{2}}{k_{Fn}^{2}+k_{1}^{2}}\frac{(k_{Fn}-k_{2})^{2}}{(k_{Fn}-k_{2})^{2}+k_{3}^{2}}\;,
\end{equation}
in which $\Theta(x)$ denotes the Heaviside unit step function. The values of the parameters $\Delta_0$, $k_{max}$, $k_1$, $k_2$, $k_3$ are given 
in Tab.~\ref{table:fit:gap}. 
As shown in Fig.~\ref{gap:pnm:dddi:A}, expression~(\ref{fit:eq}) can well reproduce the gap by  \cite{Cao2006a}. 

In the literature several \textsl{ab-initio} calculations of the pairing gap have been performed~\citep{gezerlis2010}.
They are included in the gray band in Fig.\ref{gap:pnm:dddi:A}. 
There are some discrepancies among them, as indicated by the thickness of the band, but the parametrization of the pairing gap that we consider (solid green line) remains within this band.

\begin{table}%\footnotesize
\caption{Values of the parameters entering into Eq.~(\ref{fit:eq}).}
\setlength{\tabcolsep}{.12in}
\renewcommand{\arraystretch}{1.6}
\begin{center}
\begin{tabular}{ccccc}
\hline
$k_{max}$ [fm$^{-1}$]& $\Delta_{0}$ [MeV] & $k_{1}$ [fm$^{-1}$]& $k_{2}$[fm$^{-1}$] & $k_{3}$[fm$^{-1}$]\\
\hline
1.37 &3.37968 & 0.556092 &1.38236& 0.327517\\
\hline
\end{tabular}
\label{table:fit:gap}
\end{center}
\end{table}

\begin{figure}
\begin{center}
\includegraphics[width=0.38\textwidth,angle=-90]{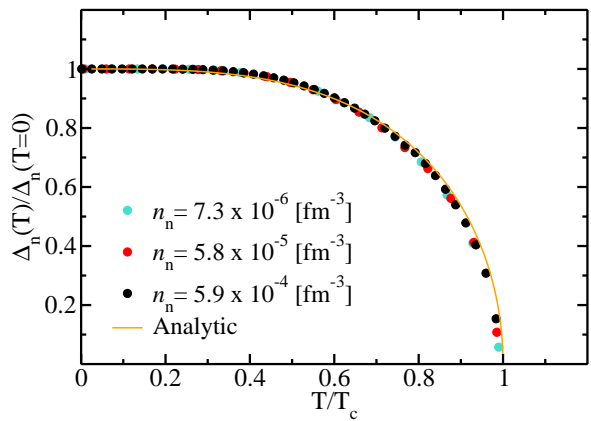}
\end{center}
\caption{(Color online) Evolution of the $^1\text{S}_{0}$ neutron pairing gap as a function of temperature for some representative value of the density of the system 
expressed in $[\text{fm}^{-3}]$. The dots represent the exact FTHFB results, while the solid line is obtained using Eq.~(\ref{delta:fit}).}
\label{gap:evoT}
\end{figure}

Injecting the gap expression~(\ref{fit:eq}) into the definition of the effective pairing force~(\ref{func:bsk16}), we have solved the FTHFB 
Eqs.~(\ref{bcs:pnm1})-(\ref{bcs:pnm2}) and we have determined the evolution of the pairing gaps as a function of temperature. 
The value of the pairing cutoff can be arbitrarily chosen. 
As shown in Fig.\ref{gap:evoT}, the gap decreases with increasing temperature, 
and eventually vanishes at the critical temperature $T_{sf}$. The temperature dependence of the pairing gap can be accurately represented by the following expression~\citep{Goriely1996}

\begin{equation}\label{delta:fit}
\Delta_n(T)=\Delta_n(T=0)\sqrt{1-\left(\frac{T}{T_{sf}} \right)^{3.23}}\Theta(T-T_{sf})\, ,
\end{equation}
for $T\leq T_{sf}$. 
The critical temperature can be obtained from the pairing gap at zero temperature from the well-known relation~\citep{Book:LL2}
\begin{equation}\label{eq:Tc}
 T_{sf}=\frac{\exp(\xi)}{\pi} \Delta_n(T=0)\, ,
\end{equation}
where $\xi\approx 0.577$ is the Euler-Mascheroni constant. Equations~(\ref{fit:eq}), (\ref{delta:fit}) and (\ref{eq:Tc}) provide a convenient 
parametrization of the evolution the pairing gaps as a function of both temperature and density. 

\begin{figure}
\begin{center}
\includegraphics[width=0.43\textwidth,angle=-90]{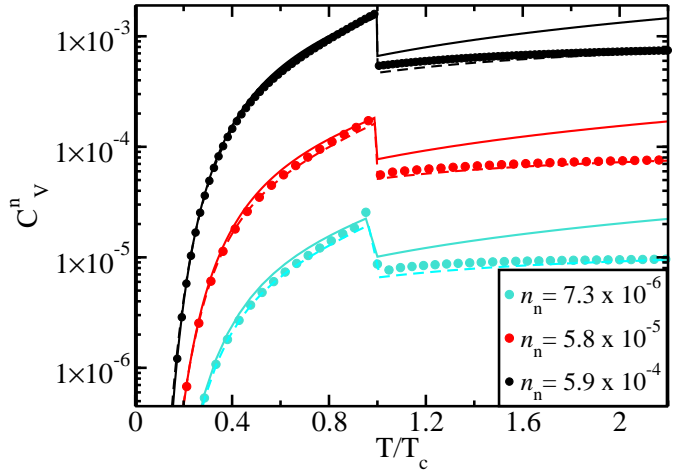}
\end{center}
\caption{(Color online) Neutron heat capacity: FTHFB results (dots) are compared to the { original expression \citet{Levenfish1994}, see Eq.(\ref{eq:levenfish}), (solid line) 
and to the new expression (\ref{cv:exp}) with refitted parameters given in Tab.\ref{tab:fit} (dashed line)}. 
The values of the density are expressed in $[\text{fm}^{-3}]$.}
\label{cv:levenfish}
\end{figure}

From the solutions of the FTHFB Eqs.~(\ref{bcs:pnm1})-(\ref{bcs:pnm2}), we have calculated the heat capacity using Eq.~(\ref{eq:cv}) and 
compared the results with those obtained using the analytical formula proposed by \citet{Levenfish1994}. As shown in Fig. \ref{cv:levenfish}, 
this formula reproduces very well the { FTHFB results in all cases and for the domain of temperature ranging from 0 to few MeV.}

\begin{table}%\footnotesize
\setlength{\tabcolsep}{.15in}
\renewcommand{\arraystretch}{1.6}
\begin{center}
\begin{tabular}{c|cc}
\hline
 & LY & present\\
\hline
$a_{0}$ &0.4186&-0.7294\\
$a_{1}$&1.007&2.331\\
$a_{2}$&0.501&0.806\\
$b_{0}$&1.456&0.743\\
$b_{1}$ &1.456&0.743\\
$\gamma$ &2.5&2.5\\
$c_{0}$ &1.456&-2.689\\
$c_{1}$ &0.157&-3.800\\
$c_{2}$& 1.764& 0.826\\
\hline
\hline
\end{tabular}
\caption{Values of the parameters appearing in Eq.(\ref{eq:r00}). LY refers to the original values given by \citet{Levenfish1994}.}
\label{tab:fit}
\end{center}
\end{table}

The formula of \citet{Levenfish1994} can be further improved by readjusting the numerical parameters of the $R_{00}$ form factor.
As shown in Fig.\ref{cv:levenfish}, the new formula for the heat capacity 
\begin{equation}
C_{V}^{n}(n_n,T)\approx\frac{3}{2}n_{n}R_{00}(T)\biggl[ 1-\exp\left(-\frac{T}{T_{cl}}\right)\biggr]
\label{eq:general}
\end{equation}
provides a very food fit to the exact FTHFB results both in the quantal and classical regimes.

%%%%%%%%%%%%%%%%%%%%%%%%%%%%%%%%%%%%%%%%%%%%%%%%%%%%%%%%
%%%    Conclusions
%%%%%%%%%%%%%%%%%%%%%%%%%%%%%%%%%%%%%%%%%%%%%%%%%%%%%%%%
\section{Conclusions}\label{sec:conclusion}

The heat capacity of dilute neutron matter has been calculated in the framework of the finite-temperature Hartree-Fock-Bogoliubov method using 
an energy density functional fitted to microscopic calculations at zero temperature. Although the heat capacity is found to be well reproduced 
by the empirical expressions of \cite{Levenfish1994} at high enough density, it differs substantially in the dilute limit corresponding to the 
shallow region of the inner crust of a neutron star. In particular, the use of Eq.(\ref{eq:levenfish}) overestimates the heat capacity of 
nonsuperfluid neutrons, hence also the thermal relaxation time of neutron-star crusts. Indeed, nonsuperfluid neutrons provide the main contribution 
to the heat capacity of the neutron-star inner crust~\citep{Gnedin2001}. We have proposed a new analytical expression obtained by combining 
Eqs.~(\ref{eq:levenfish}), (\ref{eq:r00}), (\ref{cv:exp}), that tends to the correct limits both at $T=0$ and at $T\rightarrow +\infty$. 
This expression also takes into account the superfluid transition at $T=T_{sf}$, the critical temperature $T_{sf}$ being given by Eqs.~(\ref{fit:eq}) and 
(\ref{eq:Tc}).

The presence of neutron-proton clusters in the inner crust of a neutron star may impact the heat capacity of the neutron liquid~\citep{Pizzochero2002, 
Sandulescu2004b, Monrozeau2007, Fortin2010, Chamel2010b, Pastore2012}, an effect which we haven't considered in this study. While this approximation was shown to be very 
accurate in the normal phase~\citep{Chamel2009}, it may induce larger errors in the superfluid phase. In most regions of the inner crust, the pairing gap 
hence also the critical temperature are reduced as compared to pure neutron matter calculations, but this effect can be easily included in the present 
expression for $C_V^{n}$ by simply renormalizing $T_{sf}$~\citep[see, e.g., ][]{Chamel2010b}. As a matter of fact, this reduction of $T_{sf}$ lies within the 
theoretical uncertainties of microscopic calculations in pure neutron matter. On the other hand, spatial inhomogeneities can drastically change the phase 
diagram of the neutron liquid near the neutron-drip transition~\citep{Margueron2012,Pastore2012}. 
 The effect of non-uniform matter could be parameterized as an additional correction to the uniform matter expressions,
as performed by \cite{Fortin2010}. The simple expression proposed in this work, valid in all regimes from classical's to superfluid's ones,
could efficiently be used to determine accurately the aforementioned corrections.
%{\color{blue}{ Finally, it is worth mentioning that the coupling constant of the effective interaction could be in principle present a temperature dependence, since some correlation process as density fluctuations~\citep{chamel13}, which are implicitly  included in the effective coupling constant, do have a temperature dependence; we refer to~\cite{baym2008landau} for a more detailed discussion. }}
Further studies are therefore needed to determine the 
neutron heat capacity in these layers of the star, but the present work provides an improved reference for more refined calculations 
taking into account the presence of nuclear clusters in the crust.

\section*{Acknowledgments}
This work was partially supported by the COST Action MP1304 and by the ANR SN2NS.
N. C. acknowledges support from FNRS (Belgium). A.P. thanks M. Urban for the interesting comments that motivated this work.

\bibliographystyle{mn2e}
\bibliography{biblio}

\end{document}